\documentclass[11pt]{article}

\usepackage[margin=1in]{geometry}
\usepackage{graphicx}
\usepackage{setspace}
\usepackage{titlesec}
\usepackage{fancyhdr}
\usepackage{multicol}
\usepackage{url}
\usepackage[hidelinks]{hyperref}
\usepackage{caption}
\usepackage{tcolorbox} 
\usepackage{wrapfig}
\usepackage{ragged2e}
\usepackage[backend=biber,
            style=authoryear,
            natbib=true,
            maxnames=3,
            minnames=3,
            maxcitenames=3,
            mincitenames=1,
            doi=false,
            sorting=nyt,    
            url=false,
            isbn=false,
            giveninits=true,
            eprint=false]{biblatex}
\defbibenvironment{inlinebib}
  {\noindent References:\space}
  {\unskip}
  {\printtext[labelnumberwidth]{##1}\addcomma\space}    

\AtEveryBibitem{%
  \clearfield{month}%
  \clearfield{day}%
  \clearfield{title}
  \clearfield{subtitle}%
  \clearfield{note}%
  \clearlist{language}%
}

\titleformat{\section}{\large\bfseries}{\thesection}{1em}{}
\titleformat{\subsection}{\normalsize\bfseries}{\thesubsection}{1em}{}
\titlespacing*{\section}{0pt}{1.2ex plus 0.5ex minus 0.2ex}{0.8ex}
\titlespacing*{\subsection}{0pt}{1.0ex plus 0.3ex minus 0.2ex}{0.6ex}

\title{\Huge \textbf{Extragalactic archaeology through high-resolution integrated-light spectroscopy of globular clusters}}
\author{
\large M. A. Beasley\textsuperscript{1, 2, 3}, K. Fahrion\textsuperscript{4}, A. Gvozdenko\textsuperscript{5}, S. Larsen\textsuperscript{6}\\[0.5em]
\small
\textsuperscript{1}Instituto de Astrofísica de Canarias, La Laguna, Spain,\\ 
\small
\textsuperscript{2}Universidad de La Laguna, La Laguna, Spain,\\ 
\small
\textsuperscript{3}Centre for Astrophysics and Supercomputing, Swinburne University, Australia\\ 
\small
\textsuperscript{4}Department of Astrophysics, University of Vienna, Austria\\
\small
\textsuperscript{5}{Department of Physics, Centre for Extragalactic Astronomy, Durham University, UK}\\
\small
\textsuperscript{6}{Department of Astrophysics/IMAPP, Radboud University, The Netherlands}\\[0.5em]
}
\date{\today}

\begin{document}

\maketitle
\thispagestyle{empty}

\newpage

\begin{center}
\begin{tcolorbox}[colback=blue!5!white,colframe=black!75!black, width=0.90\textwidth]

{\em  \noindent In this white paper for ESO's Expanding Horizons initiative we propose a large-scale spectroscopic facility to obtain high spectral resolution (R $\sim$ 20,000) spectroscopy for a significant fraction of all globular clusters in the nearby Universe. This will facilitate the reconstruction of galaxy assembly histories via chemical tagging, trace dark matter haloes, and measure extragalactic distances.
}

\end{tcolorbox}
\end{center}

\section{Context}
Globular clusters (GCs) are found in all massive galaxies. The idea that GCs trace the hierarchical growth of galaxies via mergers and accretion goes back many decades and is now well established through observations of GCs in the Milky Way and extragalactic systems \citep[e.g.,][]{SearleZinn1978, Harris1991, BrodieStrader2006}. As bright point-like tracers, GCs have been observed in the haloes of distant galaxies and the upcoming wide field imaging campaigns of the next 5 - 15 years with facilities such as Euclid, Roman, or LSST Rubin are expected to unveil the GC systems of most nearby galaxies. However, to unlock the potential of GCs as tracers of galaxy assembly, spectroscopy is required. Here, we propose a large-scale effort to obtain high-resolution (R $\sim$ 20,000) spectroscopy of a significant fraction of all GCs in the nearby Universe to reconstruct galaxy assembly histories, trace dark matter haloes, and measure distances.

\section{Extragalactic archaeology with globular clusters}

The general picture where GCs can trace the growth of galaxies via mergers and accretions goes back at least to the influential work of \citet{SearleZinn1978}. They observed that the metallicity distribution and lack of a radial metallicity gradient among the Milky Way outer halo GCs could not be explained by a single, rapid collapse (as proposed by \cite{Eggen1962}). Instead, they suggested that the Galactic halo was built up hierarchically through the accretion of smaller, protogalactic fragments, each potentially hosting its own population of GCs. 

\begin{wrapfigure}{r}{0.34\textwidth}
\vspace{-1cm}
  \includegraphics[width=0.34\textwidth]{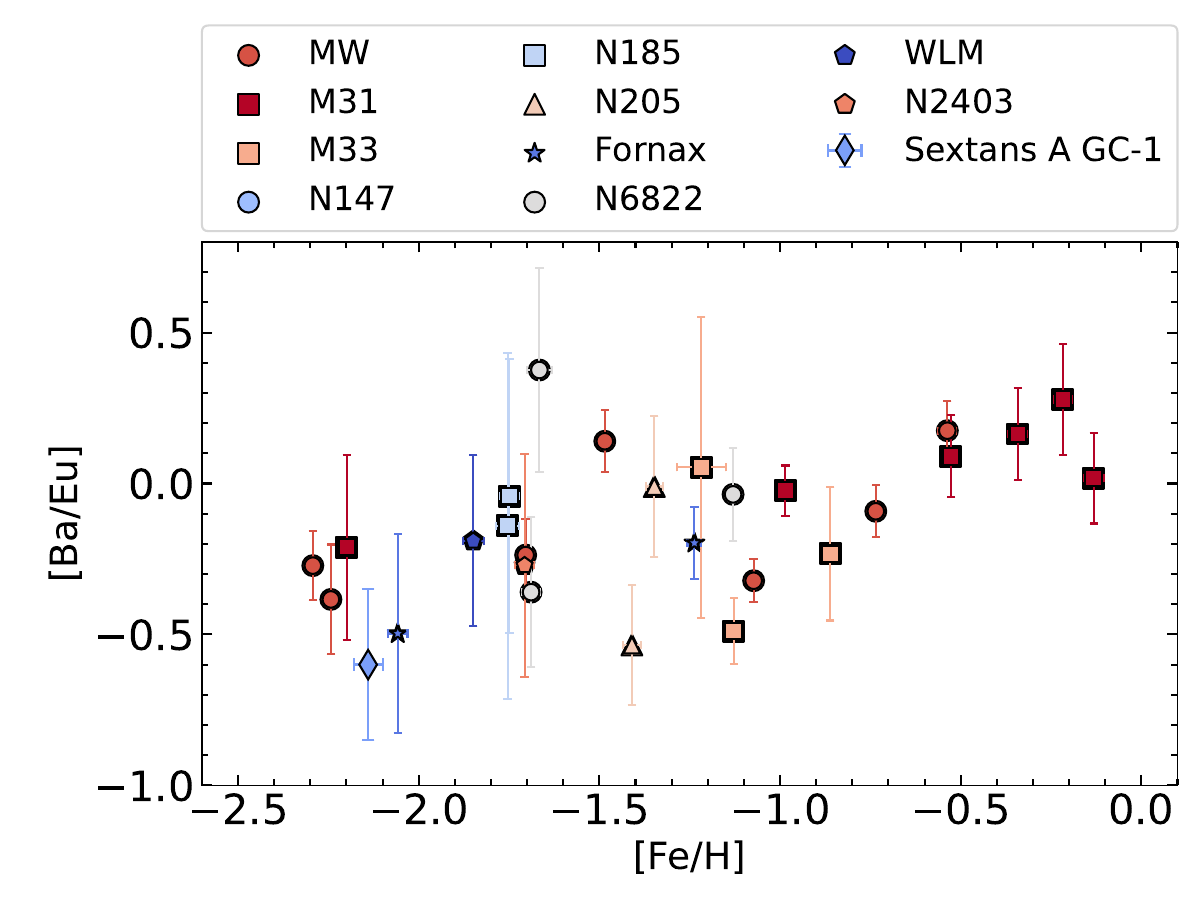}
  \vspace{-0.8cm}
  \caption{\small \textit{[Ba/Eu] versus [Fe/H] for the Local Group GCs (data from \cite{Larsen2022, Gvozdenko2024}).}}
  \vspace{-0.3cm}
\end{wrapfigure}

The \citet{SearleZinn1978} study presaged the now standard picture of hierarchical galaxy formation within $\Lambda$CDM cosmology. Subsequent decades of research have provided strong support for this model. Chemodynamical tagging of GCs using spectroscopy and Gaia astrometry has revealed that many halo GCs share common orbital and abundance signatures, linking them to accreted dwarf galaxies. Distinct merger events such as Gaia-Enceladus/Sausage, Sequoia, Kraken, and Koala have been linked to specific groups of GCs, revealing the Galaxy’s assembly history (\cite{Myeong2019, Callingham2022}).  

Specific elemental abundance patterns have been identified as potential separators between these in- and ex-situ GCs (e.g., \cite{Belokurov2023, Monty2024}). For example, by separating stars and GCs in energy-angular momentum space, \citet{Monty2024} showed that accreted Galactic GCs tend to occupy a high [Eu/Si] region, while in-situ populations exhibit notably lower ratios at a given [Fe/H]. Our ability to measure detailed chemical abundances from high-resolution integrated light (IL) spectra of GCs has advanced to the point where we can potentially perform similar work in extragalactic systems, (e.g., \cite{Sakari2015, Larsen2022, Gvozdenko2024}){\bf (Fig.~1)}. With sufficient spectral resolution (R$\sim20,000$) and signal-to-noise (>100 \AA$^{-1}$),
many chemical abundances can be measured from optical or infrared spectra in extragalactic GCs (including: Fe, Na, Mg, Si, Ca, Al, Sc, Ti, Cr, Mn, Ni, Cu, Zn, Zr, Ba, and Eu).

\section{Globular clusters as chemodynamical tracers}
The same high-resolution spectra can be used to obtain very precise line-of-sight velocities (uncertainties of order 1 km s$^{-1}$ e.g., \citealt{Fahrion2025}). This precision, that can only be achieved with high resolution spectra, allows one to identify dynamically coherent substructures such as GCs accreted in the same event. Additionally, GC velocities are used as inputs to dynamical models that constrain dark matter profiles, enclosed masses, and probe the kinematics of the host galaxy and its halo (e.g. \cite{Versic2024}). While such models exist for a handful of local galaxies, only larger samples will open the possibility to self-consistently test the relation between GC numbers/masses and dark matter halo masses, a relation which gained interest within the community in recent years due to its potential of using GC numbers as tracers of dark matter (e.g. \cite{SpitlerForbes2009}).  

\section{Extragalactic distance measurements with globular clusters}
\begin{wrapfigure}{r}{0.34\textwidth}
\vspace{-0.5cm}
  \includegraphics[width=0.34\textwidth]{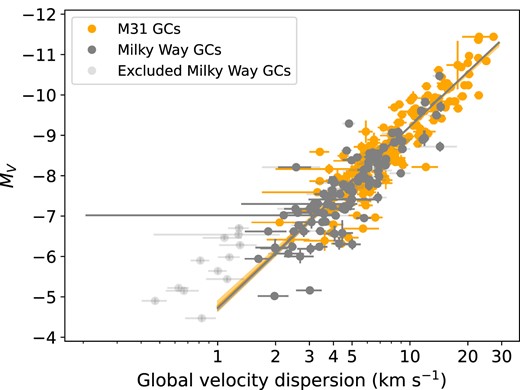}
  \vspace{-0.8cm}
  \caption{\small \textit{Relation between globular cluster velocity dispersions $\sigma$ and absolute magnitudes for Milky Way (grey) and M31 GCs (orange). GCs with unreliable $\sigma$ measurements were excluded (from \cite{Beasley2024}).}}
  \vspace{-0.3cm}
\end{wrapfigure}
Distance measurements are fundamental to all aspects of astronomy and are the cornerstone of observational cosmology. The still unresolved Hubble tension highlights the need for new distance measurement tools. We propose to use a newly developed method based on the velocity dispersions of GCs which can be obtained with high-resolution spectroscopy. This method is based on the relation between intrinsic velocity dispersion and brightness of GC, and it has been shown that this approach allows distance measurements with $3-5\%$ uncertainty, similar to other precise methods \citep{Beasley2024} {\bf (Fig. 2)}.
Only with high resolution spectroscopy, internal GC stellar velocity dispersions can be measured ($\sigma >4$ km s$^{-1}$, R$\sim$20,000 required), which will allow us to provide accurate distance measurements to galaxies within $\sim 50$\,Mpc. Connecting those distances with known supernova hosts could potentially open a new way of constraining $H_0$. Additionally, it will allow us to trace out the large-scale structure in the nearby Universe, for example, by deriving relative distances of galaxies within the same group or cluster.

\section{Technical requirements and synergies}
To exploit GCs as tracers of galaxy assembly in the nearby Universe, a high-resolution, multi-object spectrograph on a large diameter telescope is required. The facility should provide a few hundred deployable fibres at resolving power 
$R\sim20,000$, sufficient to measure detailed chemical abundances and derive line-of-sight velocities and velocity dispersions with precision of $<$1 km s$^{-1}$. A broad optical wavelength range (e.g. 4000--9000 \AA) is needed to cover key diagnostic lines, 
including r-process Eu, s-process Ba, and multiple $\alpha$-elements (Mg, Si, Ca, Ti), while distinct wavelength windows (optical, CaT) would maximise flexibility for different science cases and observing conditions. 
A field-of-view of order 10 arcminutes diameter is highly desirable to encompass the extended GC systems and stellar haloes of nearby galaxies, and a mirror size of $\sim 15$m would provide the sensitivity to reach the bright half of the GC luminosity function to at least 20 Mpc.
Compared to current facilities such as MOONS and 4MOST, a larger mirror diameter with higher spectral resolution is essential to reach the required depth, thereby fulfilling the science proposed here, and to map statistically significant samples of GCs distributed in galaxy haloes with reasonable efficiency. 

The proposed facility will fully leverage deep, wide-field optical and near-infrared imaging surveys that will have identified rich GC systems, satellite populations, and other halo tracers around nearby galaxies. High-resolution spectroscopy will turn these catalogues into comprehensive chemo-dynamical maps of galaxy assembly. The high multiplex also enables efficient use of spare fibres, which can be allocated to complementary tracers in galaxy haloes such as planetary nebulae for independent kinematic measurements, dwarf satellites for extended dynamical constraints, or background quasars to probe the circumgalactic medium through absorption-line spectroscopy. 

{
\small
\singlespacing
\printbibliography

@ARTICLE{Eggen1962,
       author = {{Eggen}, O.~J. and {Lynden-Bell}, D. and {Sandage}, A.~R.},
        title = "{Evidence from the motions of old stars that the Galaxy collapsed.}",
      journal = {\apj},
         year = 1962,
        month = nov,
       volume = {136},
        pages = {748},
          doi = {10.1086/147433},
       adsurl = {https://ui.adsabs.harvard.edu/abs/1962ApJ...136..748E}
}

@ARTICLE{Larsen2022,
       author = {{Larsen}, S.~S. and {Eitner}, P. and {Magg}, E. and {Bergemann}, M. and {Moltzer}, C.~A.~S. and {Brodie}, J.~P. and {Romanowsky}, A.~J. and {Strader}, J.},
        title = "{The chemical composition of globular clusters in the Local Group}",
      journal = {\aap},
         year = 2022,
        month = apr,
       volume = {660},
          eid = {A88},
        pages = {A88},
          doi = {10.1051/0004-6361/202142243},
archivePrefix = {arXiv},
       eprint = {2112.00081},
 primaryClass = {astro-ph.GA},
       adsurl = {https://ui.adsabs.harvard.edu/abs/2022A&A...660A..88L}
}

@ARTICLE{Sakari2015,
       author = {{Sakari}, Charli M. and {Venn}, Kim A. and {Mackey}, Dougal and {Shetrone}, Matthew D. and {Dotter}, Aaron and {Ferguson}, Annette M.~N. and {Huxor}, Avon},
        title = "{Integrated light chemical tagging analyses of seven M31 outer halo globular clusters from the Pan-Andromeda Archaeological Survey}",
      journal = {\mnras},
         year = 2015,
        month = apr,
       volume = {448},
       number = {2},
        pages = {1314-1334},
          doi = {10.1093/mnras/stv020},
archivePrefix = {arXiv},
       eprint = {1501.04626},
 primaryClass = {astro-ph.GA},
       adsurl = {https://ui.adsabs.harvard.edu/abs/2015MNRAS.448.1314S}
}

@ARTICLE{Fahrion2025,
       author = {{Fahrion}, Katja and {Beasley}, Michael A. and {Emsellem}, Eric and {Gvozdenko}, Anastasia and {M{\"u}ller}, Oliver and {Rejkuba}, Marina and {Chies-Santos}, Ana L.},
        title = "{Globular clusters in M 104: Tracing kinematics and metallicities from the centre to the halo}",
      journal = {\aap},
         year = 2025,
        month = oct,
       volume = {702},
          eid = {A59},
        pages = {A59},
          doi = {10.1051/0004-6361/202555465},
archivePrefix = {arXiv},
       eprint = {2508.10100},
 primaryClass = {astro-ph.GA},
       adsurl = {https://ui.adsabs.harvard.edu/abs/2025A&A...702A..59F}
}

@ARTICLE{Beasley2024,
       author = {{Beasley}, Michael A. and {Fahrion}, Katja and {Gvozdenko}, Anastasia},
        title = "{Measuring distances to galaxies with globular cluster velocity dispersions}",
      journal = {\mnras},
         year = 2024,
        month = jan,
       volume = {527},
       number = {3},
        pages = {5767-5775},
          doi = {10.1093/mnras/stad3541},
archivePrefix = {arXiv},
       eprint = {2312.01420},
 primaryClass = {astro-ph.GA},
       adsurl = {https://ui.adsabs.harvard.edu/abs/2024MNRAS.527.5767B}
}

@ARTICLE{SpitlerForbes2009,
       author = {{Spitler}, L.~R. and {Forbes}, D.~A.},
        title = "{A new method for estimating dark matter halo masses using globular cluster systems}",
      journal = {\mnras},
         year = 2009,
        month = jan,
       volume = {392},
       number = {1},
        pages = {L1-L5},
          doi = {10.1111/j.1745-3933.2008.00567.x},
archivePrefix = {arXiv},
       eprint = {0809.5057},
 primaryClass = {astro-ph},
       adsurl = {https://ui.adsabs.harvard.edu/abs/2009MNRAS.392L...1S}
}

@ARTICLE{Versic2024,
       author = {{Ver{\v{s}}i{\v{c}}}, Tadeja and {Thater}, Sabine and {van de Ven}, Glenn and {Watkins}, Laura L. and {Jethwa}, Prashin and {Leaman}, Ryan and {Zocchi}, Alice},
        title = "{Total mass slopes and enclosed mass constrained by globular cluster system dynamics}",
      journal = {\aap},
         year = 2024,
        month = jan,
       volume = {681},
          eid = {A46},
        pages = {A46},
          doi = {10.1051/0004-6361/202347413},
archivePrefix = {arXiv},
       eprint = {2310.12338},
 primaryClass = {astro-ph.GA},
       adsurl = {https://ui.adsabs.harvard.edu/abs/2024A&A...681A..46V}
}

@ARTICLE{Monty2024,
       author = {{Monty}, Stephanie and {Belokurov}, Vasily and {Sanders}, Jason L. and {Hansen}, Terese T. and {Sakari}, Charli M. and {McKenzie}, Madeleine and {Myeong}, GyuChul and {Davies}, Elliot Y. and {Ardern-Arentsen}, Anke and {Massari}, Davide},
        title = "{The ratio of [Eu/{\ensuremath{\alpha}}] differentiates accreted/in situ Milky Way stars across metallicities, as indicated by both field stars and globular clusters}",
      journal = {\mnras},
         year = 2024,
        month = sep,
       volume = {533},
       number = {2},
        pages = {2420-2440},
          doi = {10.1093/mnras/stae1895},
archivePrefix = {arXiv},
       eprint = {2405.08963},
 primaryClass = {astro-ph.GA},
       adsurl = {https://ui.adsabs.harvard.edu/abs/2024MNRAS.533.2420M}
}

@ARTICLE{Gvozdenko2024,
       author = {{Gvozdenko}, A. and {Larsen}, S.~S. and {Beasley}, M.~A. and {Cabrera-Ziri}, I. and {Eitner}, P. and {Battaglia}, G. and {Leaman}, R.},
        title = "{Detailed chemical composition of the globular cluster Sextans A GC-1 on the outskirts of the Local Group}",
      journal = {\aap},
         year = 2024,
        month = may,
       volume = {685},
          eid = {A154},
        pages = {A154},
          doi = {10.1051/0004-6361/202346859},
archivePrefix = {arXiv},
       eprint = {2403.10597},
 primaryClass = {astro-ph.GA},
       adsurl = {https://ui.adsabs.harvard.edu/abs/2024A&A...685A.154G}
}

@ARTICLE{Belokurov2023,
       author = {{Belokurov}, Vasily and {Kravtsov}, Andrey},
        title = "{Nitrogen enrichment and clustered star formation at the dawn of the Galaxy}",
      journal = {\mnras},
         year = 2023,
        month = nov,
       volume = {525},
       number = {3},
        pages = {4456-4473},
          doi = {10.1093/mnras/stad2241},
archivePrefix = {arXiv},
       eprint = {2306.00060},
 primaryClass = {astro-ph.GA},
       adsurl = {https://ui.adsabs.harvard.edu/abs/2023MNRAS.525.4456B}
}

@ARTICLE{BrodieStrader2006,
       author = {{Brodie}, Jean P. and {Strader}, Jay},
        title = "{Extragalactic Globular Clusters and Galaxy Formation}",
      journal = {\araa},
         year = 2006,
        month = sep,
       volume = {44},
       number = {1},
        pages = {193-267},
          doi = {10.1146/annurev.astro.44.051905.092441},
archivePrefix = {arXiv},
       eprint = {astro-ph/0602601},
 primaryClass = {astro-ph},
       adsurl = {https://ui.adsabs.harvard.edu/abs/2006ARA&A..44..193B}
}

@ARTICLE{Myeong2019,
       author = {{Myeong}, G.~C. and {Vasiliev}, E. and {Iorio}, G. and {Evans}, N.~W. and {Belokurov}, V.},
        title = "{Evidence for two early accretion events that built the Milky Way stellar halo}",
      journal = {\mnras},
         year = 2019,
        month = sep,
       volume = {488},
       number = {1},
        pages = {1235-1247},
          doi = {10.1093/mnras/stz1770},
archivePrefix = {arXiv},
       eprint = {1904.03185},
 primaryClass = {astro-ph.GA},
       adsurl = {https://ui.adsabs.harvard.edu/abs/2019MNRAS.488.1235M}
}

@ARTICLE{Callingham2022,
       author = {{Callingham}, Thomas M. and {Cautun}, Marius and {Deason}, Alis J. and {Frenk}, Carlos S. and {Grand}, Robert J.~J. and {Marinacci}, Federico},
        title = "{The chemo-dynamical groups of Galactic globular clusters}",
      journal = {\mnras},
         year = 2022,
        month = jul,
       volume = {513},
       number = {3},
        pages = {4107-4129},
          doi = {10.1093/mnras/stac1145},
archivePrefix = {arXiv},
       eprint = {2202.00591},
 primaryClass = {astro-ph.GA},
       adsurl = {https://ui.adsabs.harvard.edu/abs/2022MNRAS.513.4107C}
}

@ARTICLE{Harris1991,
       author = {{Harris}, William E.},
        title = "{Globular cluster systems in galaxies beyond the Local Group.}",
      journal = {\araa},
         year = 1991,
        month = jan,
       volume = {29},
        pages = {543-579},
          doi = {10.1146/annurev.aa.29.090191.002551},
       adsurl = {https://ui.adsabs.harvard.edu/abs/1991ARA&A..29..543H}
}

@ARTICLE{SearleZinn1978,
       author = {{Searle}, L. and {Zinn}, R.},
        title = "{Composition of halo clusters and the formation of the galactic halo.}",
      journal = {\apj},
         year = 1978,
        month = oct,
       volume = {225},
        pages = {357-379},
          doi = {10.1086/156499},
       adsurl = {https://ui.adsabs.harvard.edu/abs/1978ApJ...225..357S}
}
}

\end{document}